# Design issues for distributed mobile social networks

**D. N. Kallergis[1], I.E. Foukarakis[2], G. N. Prezerakos[1]**

[1] Dpt. of Electronic Computer Systems, Technological Educational Institute (TEI) of Piraeus, Aigaleo, Greece, E-mail: {d.kallergis, prezerak}@teipir.gr

[2] Dpt. of Telecommunications Science and Technology, University of Peloponnese, Tripoli, Greece, E-mail: ifouk@uop.gr

### ABSTRACT

Social networks and their applications have become extremely popular during the last years, mostly targeting users via the web. However, it has been recently observed an interest to offer social network services to mobile users. Telecom operators attempt to integrate existing social networks to their systems or develop new ones, in order to offer new services to their subscribers. Subsequently, emphasis is given to the user-context modeling, as well as to the integration of sources that leads to the summarized collection of information anchored to the user; such as its location or its mobile device type, etc.

In this paper we discuss the most important factors and challenges encountered during the design of such a system on architectural, technological and tool level.

*Keywords: user context, distributed social-networking, context-aware platform, social graphs, mobile networks.*

## 1. Introduction

Social network sites (SNSs) are to turn to mobile; this means that their former definition as introduced by D. Boyd and N. Ellison [1] and as enriched by D. Beer [2], a year later, unambiguously does not exist any longer. Famous examples of social networking sites, following this trend, are MySpace, Facebook and LinkedIn, which are designed to maintain users' social profiles, as well as the virtual social network formed among them. This network is deployed considering common interests, friends, or even professional activity, as it is declared by users. The three of them are to turn to mobile, through using a mobile version of an existing browser, in order to facilitate the user within the context of the site. In addition, a growing number of mobile specific social networks and related services has appeared in the public, enhancing the user's experience by using additional communication channels, including WAP, SMS and MMS. This growth is driven by the advances in technologies supporting social network architectures, the improved devices capabilities of modern mobile terminals and the lower prices in the data plans of the mobile operators.

In this paper, we are going to discuss the key elements in designing a mobile social network. Furthermore, we are going to attempt to map these design issues to existing mobile social networks, both available on production and research environments.

## 2. Aspects of system design

Mobile social networks can be seen as the adaptation of social networks for mobile devices. Thus, it is easily assumed that the same elements that take a hand in social network design decisions apply to mobile versions as well. However, the user's mobile device has both additional functionalities and limitations when compared to a user's personal computer or laptop, making some of the design issues more complex, or even adding new ones. In this paragraph we are going to focus on the most important of the design aspects for mobile social networking.

2.1 User Context

The most important advantage of a mobile social network is that it allows access to more context sources. The user interacts with the mobile social network through his/her mobile device, allowing multiple types of information to be passed to the social network, enabling improved network experience and service personalization, while enhancing his/her social presence. However, these features are not utilized by most of the existing social networks [3].

The most important types of user context include:

*A. User Profile*

A typical user profile consists of a unique user identifier, various pieces of personal information, a list of friends and a list of social groups (or networks) consisting of users sharing some common interest. It contains mostly information that does not frequently change over time.

*B. Location*

Location is possibly the most commonly discussed characteristic related to user context. Many different technologies exist to allow retrieving the user's location based on his/her mobile device, with most of them being discussed in [5]. The user's location can be denoted through using different levels of abstraction, varying from exact position, using a coordinate system, to proximity based systems [6], or even more abstract ones, like a place name; i.e. Eiffel Tower, Paris.

Nowadays, location data are used in connection with Location Based Services (LBS). LBS utilize the location of the mobile device to offer additional value to the mobile subscriber or to a third party (i.e. the service provider). This is already applicable to a wide variety of applications, including vehicle tracking regarding parcel delivery, discovering the nearest working bank branch, direct advertising based on the user's current location (i.e. an airport's duty free stores area), providing location-based games in an amusement park, or even providing auto insurance policies for the user's current activity plan.

*C. Identity*

While accessing different services, a user might use different credentials, in order to authenticate him/her. The information attached to each identity can be correlated with the user's context, in order to identify his/her current status (e.g. working, travelling, etc.). Apart from typing the user's login name/password, mobile devices offer alternative authentication methods, including the user's MSISDN, submitting information over Bluetooth or displaying barcodes or other visual codes on the device's screen.

*D. Device type and information*

The user's device type maybe plays the most important role within every possible scenario. The range within each one operates, as well as the capability to participate in more than one network simultaneously gives a different perspective in terms of how the user's context is going to be extracted. All the user context types mentioned above are largely affected by the user's mobile device capabilities.

Nowadays, most of the advanced mobile terminals (including PDAs and smartphones) come equipped with multiple wireless interfaces, including Bluetooth, Wi-Fi, assisted GPS (A-GPS) and cellular radio, allowing the devices to communicate and more than this; discover similar devices in the periphery. When we talk about outdoor use, GSM and A-GPS work efficiently, whilst indoor use preferably refers to Wi-Fi, Bluetooth, RFID, Ultra-wideband (UWB) and ZigBee.

Several parameters interact, in order to choose the most efficient technology and, thus, the most preferable device. We address some of them, since we believe that they maintain the most crucial part, even in a mobile social network scenario: {a} accuracy (during positioning), {b} data rate {c} encryption {d} indoor/outdoor use {e} latency {f} possible interference {g} signal range {h} scale of network deployment.

2.2 Social network models and graphs

In the core of every social network the social graph exists. This component is the model of the relationships among users and/or other components of the social network (e.g. applications). The social graph can be considered as a regular graph, where each node represents an entity in the social network (user, application, etc.), and every edge represents a relationship between the two entities. Depending on the model of the social network, various types of graphs can be utilized. If the case is that each entity must accept the relationship as a two-part relationship (mutual friend model), an undirected graph can be used, while in cases where a user watches a friend, a directed one can be used. Weighted graphs can be used to model the importance of a relationship.

The above model is pretty simplistic, as the network size affects largely the architecture of the social graph. Social networks with a small number of users can be stored in a relatively small graph, managed in a single server. However, as the size increases, different architectures may be used to allow storing and traversing the graph, utilizing different technologies. The size of the network also plays an important role in the decision, if the actual network's structure will exist in a centralized server/server farm, or will be distributed across multiple nodes that can be operated by different entities.

2.3 Application or platform

Another important option is whether the mobile social network will be an application or it will allow integration with third party applications. A standalone application will require a user interface, in order for the members of the social network to navigate and perform their tasks, possibly limiting the user's actions to specific tasks as well.

A more open approach is to design the mobile social network allowing third party developers to integrate their applications. This can be achieved either by offering functionality through public APIs or by scripting languages. Although this approach enables new applications to run inside the mobile SN, it requires additional effort in design, security and privacy.

2.4 New social network or integration with existing

The simplest way to offer a social network to mobile subscribers is by integrating with an existing one. A subset of the services offered by the social network or adapted versions can become available to the users. This approach allows the user to enhance his experience by accessing his already existing account from his mobile device. However, this approach makes it difficult to utilize user context.

On the other hand, the design and development of a new SN allows integrating mobile specific technologies. The greater cost is that the new SN will have a small number of users in the beginning, making it necessary to advertise its existence and convince the users to join in.

2.5 User interface

One of the most important issues for mobile social networks is their user interface. At present, most social networks are accessible through the user's web browser. However, in the mobile world, a variety of different web browsers exists, as well as device configurations. Screen size, colours and network bandwidth affect the final user interface largely, making content adaptation a major requirement. The user interface's designer may choose to support a selected subset of the available devices, e.g. selecting only the high-end devices. Another option is to develop custom applications, as the GUI, to the mobile social network. The different device operating systems (Symbian, Android, Windows Mobile, etc.) raise the difficulty in developing a universal application that will support all devices, but platform-specific applications offer improved user interface experience.

2.6 Target audience

From the operators' point of view, a mobile social network is a great opportunity to provide value added services to customers. A social network can be developed and used for many different purposes, including promotion of new services or devices, supporting groups of people in specific aspects of their life (work, entertainment, education, etc.), and satisfy certain needs (communication, news). The target group and scope of the social network may dictate its functional and architectural requirements.

2.7 User Generated Content

Modern mobile devices offer a wide capability of sensors that can be used to capture multimedia content. Video, sound and image files, generated on-the-fly by the user, are most likely to represent something valuable to him/her that he/she is willing to share. The same apply to text, either coming from SMS and WAP or mobile Web. An important aspect of the mobile user generated content is that some devices annotate the content by adding location and time tags.

2.8 Privacy

Privacy is one of the most discussed issues of social networks. Although different approaches have been proposed, in the case of mobile social networks, additional effort must be made. Since the user will be accessible via his/her mobile device, there is a need to protect him/her from obtrusive applications and/or members of the social network. Moreover, as he/she may continuously be connected to the social network via his/her mobile terminal, specific actions must be taken to ensure that only the data he/she has selected are available.

**3. Taxonomy of existing mobile social networks**

The number of mobile social networks or related services has grown during the last years. Until now the most widely adopted and used by the users are the mobile versions of existing social networks, including Facebook, Twitter, MySpace and LinkedIn. Note that although some of the above elements are already a part of those social networks' architecture, they do not have any mobile specific functionality. An exception is Facebook's SDK for iPhone and Android. Some other wide spread solutions, offered by mobile operators or vendors, include Timescape and Vodafone 360. These two tools are not exactly social networks, but aggregators for different popular web-based services, including the previously mentioned social networks, the user's contact list, email, communication applications etc. They can be accessed in a variety of ways, from standard Web/WAP pages to custom applications. However, all these approaches do not take advantage of the user's context, except for the a priori knowledge of the mobile device's environment.

An interesting approach on handling user context for enhancing the social presence of a user is MIT's Serendipity [7]. Serendipity keeps the users' profile saved in a central server. The users may provide weights that determine the importance of each piece of information, in order to calculate a similarity score. In accordance to this score, which is calculated when extracting the commonalities between two proximate profiles, the system alerts the two users that someone nearby might interest them; this is the case where the calculation result is above the threshold set by the users. Loopt and Aka-aki have also followed similar approaches; by allowing each user to know which of his/her friends are close.

| Mobile Social Network | URL | User Context Aware | APIs | Mobile Specific | User Interface | User Generated Content |
|---|---|---|---|---|---|---|
| Facebook | 0.facebook.com | No | SDK, REST | No | Web/WAP | Yes |
| Twitter | m.twitter.com | No | REST, Streaming | No | Web/WAP, Application | Yes |
| MySpace | m.myspace.com | No | REST | No | Web/WAP, application | Yes |
| LinkedIn | m.linkedin.com/ | No | REST | No | Web/WAP | Yes |
| Timescape | - | No | - | Yes | Application | Yes |
| Vodafone 360 | - | No | Native API | Yes | Application | Yes |
| Socialight | socialight.com | Yes | REST | Yes | Custom applications | Yes |
| Aka-aki | m.aka-aki.com | Yes | - | Yes | Application | Yes |
| Loopt | www.loopt.com | Yes | - | Yes | Application | Yes |
| Gowalla | m.gowalla.com | Yes | REST | Yes | Application, Web/WAP | Yes |
| Foursquare | foursquare.com | Yes | Web API | Yes | Application, SMS | Yes |
| MobiClique | *see* [4] | Yes | No | Yes | Application | Yes |
| Serendipity | *see* [7] | Yes | No | Yes | Application | No |

**Table 1:** Overview of Mobile Social Networks

Going further, MobiClique [4] is a networking middleware, extending the principles Serendipity has set. Based on opportunistic connections among neighboring devices (rather than relying on a central server), it uses a store-carry-forward technique, enabling the users to "advertise" their current preferences and actions; simulating a behavior of a physical community among individuals. Thus, the users are alerted and have to choose if they want to exchange; from friendship contacts to content dissemination. Moreover, their devices are set to continuously discover new devices or communities, increasing the size of the social graph.

Some of the mobile social networks that have met success include Socialight, Gowalla and Foursquare. These native mobile (location-based) social networks exclusively focus on mobile communication, location based services, augmented reality, etc. The capabilities of these networks include integration with the specialties of mobile operating systems, as well as conformation to different mobile device specifications and multiple communication channels. Their success is primarily driven by user generated content. Location-aware content generation and sharing is performed in a user friendly manner.

Finally, it is also important to note that all of existing mobile social networks examined offer a centralized approach on storing and using the user's profile and context.

The findings of our research are summarized in Table 1.

## 4. Conclusion

Mobile social networks are emerging, offering both wide variety of applications and the ground for development of next generation applications. However, there is still space for new ones, combining elements that are not yet highly utilized. User context is currently limited to location-based services, and is mostly utilized by new mobile social networks, thus creating a requirement for combining the large number of users of the existing social networks with context awareness.

The next generation of mobile social networks is extremely likely to include distributed versions of existing ones. Supporting technologies are already mature. This is also depicted in [8], which names some of the candidate technologies for implementing a distributed mobile social network on data exchange, authentication and aggregation levels.